\documentstyle[psfig]{astropart}
\renewcommand{\today}{12 August 1998}
\newcommand{\diff}{{\rm d}}

\newcommand{\sigmaT}{\sigma_{\rm T}}
\newcommand{\drange}{\rho}
\newcommand{\PSR}{PSR~B1259$-$63}
\newcommand{\PSRjb}{PSR~J0045$-$73}
\newcommand{\Rstar}{R_*}
\newcommand{\Lstar}{L_*}

\begin{document}
%
\begin{frontmatter}
\title{Inverse Compton Emission of TeV Gamma Rays from \PSR}
\author[MPIK]{J. G. Kirk}
\author[RCfTA]{Lewis Ball}
\author[RCfTA]{Olaf Skj\ae raasen\thanksref{OSLO}}
\address[MPIK]{Max-Planck-Institut f\"ur Kernphysik,
Postfach 10 39 80, D--69029, Heidelberg, Germany}
\address[RCfTA]{Research Centre for Theoretical Astrophysics,
University of Sydney, N.S.W. 2006, Australia}
\thanks[OSLO]{Institute of Theoretical Astrophysics, University
of Oslo, P.O. Box 1029 Blindern, N-0315 Oslo, Norway}

\begin{abstract}
We derive light curves for the hard $\gamma$-ray emission,
at energies up to several TeV,
expected from the unique pulsar/Be-star binary system \PSR .
This is the only known system in our galaxy in which a radio
pulsar is orbiting a main sequence star.
We show that inverse Compton emission from the electrons
and positrons in the shocked pulsar wind,
scattering target photons from the Be star,
produces a flux of hard $\gamma$-rays that should be above the
sensitivity threshold of present day atmospheric Cerenkov detectors.
Furthermore, we predict that the flux of hard $\gamma$-rays
produced via this mechanism has a characteristic variation with
orbital phase that should be observable,
and which is not expected from any other mechanism.
\end{abstract}
\end{frontmatter}

%
\section{Introduction}

Pulsars are thought to lose their rotational energy by driving a
relativistic wind of electrons, positrons and possibly ions.
The radio pulsar \PSR\ is in a binary system,
and is orbiting a luminous, massive star
[Johnston et al.\ 1996].
The orbit is highly eccentric ($e\sim 0.87$)
and at periastron the pulsar is just $23\Rstar$ from its companion,
a 10th magnitude B2e star, SS2883, of radius $\Rstar\sim 6 R_\odot$
[Johnston et al.\ 1992, 1994].

Observations indicate that this pulsar system is a source of
unpulsed X-ray emission throughout its orbit
[Hirayama et al.\ 1996].
The wind of PSR B1259$-$63 is likely to be confined by pressure balance with
the strong Be-star outflow
[Melatos, Johnston \& Melrose 1995],
and electrons and positrons will be accelerated and isotropised
at the shock which terminates the relativistic pulsar wind.
The observed X-ray emission has been interpreted as synchrotron emission
from the shocked, relativistic electrons and positrons
of the pulsar wind
[Tavani \& Arons 1997].

The high luminosity of the pulsar's companion star, together with
the relatively small separation of the binary,
raises the possibility that the inverse Compton scattering of
photons from the companion star by the relativistic electrons
and positrons from the pulsar wind may be important.
While the shocked electrons and positrons radiate synchrotron emission
at energies in the X-ray band from keV to MeV,
inverse Compton scattering of the photons from the Be star
will produce high energy photons which
are likely to be in the GeV to TeV range.

\PSR\ is the only known galactic radio pulsar which is orbiting a
main-sequence companion.
Its spin down luminosity is $8.3\times 10^{28}\;$W.
If a fraction of $10^{-3}$ or more of this luminosity were converted
into TeV photons via inverse Compton scattering, the flux
at Earth would exceed the sensitivity threshold of imaging
Cerenkov detectors such as that operated by the CANGAROO collaboration,
which is $\sim 10^{-12}\;{\rm photons \, cm^{-2} \, s^{-1}}$
[Kifune et al.\ 1995].
The only other similar binary pulsar system, \PSRjb\
[Kaspi et al.\ 1994, Bell et al.\ 1995],
is in the Small Magellanic Cloud and is the most distant pulsar known.
Although its binary separation at periastron is even smaller than
that of \PSR, it is not likely to be observable in hard $\gamma$-rays
because of its lower spin down luminosity and greater distance.

TeV $\gamma$-rays have been observed from the Crab nebula
[e.g.\ Vacanti et al.\ 1991,
HEGRA Collaboration 1996,
Tanimori et al.\ 1998],
and are attributed to inverse Compton scattering of self-produced
synchrotron photons by electrons and positrons accelerated at the
termination shock of a relativistic wind from the pulsar
[De Jager \& Harding 1992].
The emission from other pulsar nebulae that have been detected
at TeV energies
[see Kifune 1997 for a review]
is attributed to inverse Compton scattering,
by the shocked pulsar wind,
of photons from the $2.7\,$K microwave background
[Harding 1996].
None of the pulsars detected in hard $\gamma$-rays is in a binary system.

In the \PSR\ binary system, photons from the Be star form the most
important targets for inverse Compton scattering.
The effect of this process on the distribution of synchrotron emitting
electrons is included in the treatment of  
Tavani \& Arons [1997].
However, they did not discuss the resulting $\gamma$-ray emission.
In this paper we compute both the synchrotron and inverse Compton flux
emitted by the electrons and positrons accelerated at the
termination shock of the relativistic wind of \PSR ,
as a function of the orbital phase of the binary.
The highly energetic inverse Compton photons may be absorbed by
photon-photon pair creation whilst propagating through the
radiation field of the companion star, and the effective optical depth due
to this process is included in our calculations.
The parameters which describe the pulsar wind are determined
approximately from the observed X-ray fluxes near periastron,
and the model is then used to predict the light curves
at energies from 100 MeV to 1 TeV.
The modelling indicates that it may be possible to detect \PSR\
using an imaging Cerenkov detector such as that operated by the
CANGAROO collaboration.
To date, only a limited number of observations of this object have been
made using this telescope, with the possibility of at best a
marginal detection
[Sako et al.\ 1997,
Patterson \& Edwards, private communication].
We show that the inverse Compton scattering of Be-star
photons produces an unambiguous variation in the
hard $\gamma$-ray flux which should be observable,
and which is unlikely to result from any other mechanism.

\section{Emission from a shocked pulsar wind}

The wind of the Crab pulsar is thought to have a
Lorentz factor of $\Gamma\approx10^6$, and to terminate at a collisionless
shock front situated some $3\times 10^{15}\,$m from the neutron star
where it attains pressure balance with the surrounding material
[Rees \& Gunn 1974].
In the
Kennel \& Coroniti [1984]
model of the Crab nebula,
the pulsar wind electrons are accelerated at this shock front
and the kinetic energy of the wind is converted into a power-law distribution
of electrons extending over several decades in energy.
(Note that when used alone, the term `electron' can be taken
to refer to both electrons and positrons in the pulsar wind.)
The synchrotron emission from these particles is detected at all
frequencies from the radio regime to $\gamma$-rays
of energies up to $10\,$GeV.
Gamma-rays of higher energy are believed to be produced
by a synchrotron self Compton process whereby the shocked
synchrotron-emitting electrons inverse Compton scatter
synchrotron photons at infrared to optical frequencies
[De Jager \& Harding 1992].

Following
Kennel \& Coroniti [1984],
we assume that the shock front that terminates the wind of \PSR\
accelerates the electrons incident upon it and injects them into
the downstream region such that when integrated over the shock surface, the
injection rate is isotropic in the rest frame of the pulsar,
and is a power-law in momentum extending from 
a lower limit $\gamma_1$ to an upper limit $\gamma_2$.

The injection rate in an 
interval $\diff\gamma$, integrated over solid angle, is
$Q_{\rm e}(\gamma)\diff\gamma$ where
\begin{eqnarray}
Q_{\rm e}(\gamma)&=&
\left\{
\begin{array}{ll}
Q_0 \gamma^{-q} & {\rm for\ }\gamma_1<\gamma<\gamma_2 \;, \\
0                & {\rm otherwise}\;.
\end{array}
\right.
\label{espectrum}
\end{eqnarray}
Assuming that the unshocked pulsar wind contains monoenergetic electrons
and positrons of Lorentz factor $\Gamma$,
one can eliminate two of the three free parameters
$Q_0$, $\gamma_1$ and $\gamma_2$ in terms of $\Gamma$ and that 
fraction $\eta$ of the pulsar luminosity $L_{\rm p}$ 
which is channelled into relativistic
electrons: $\eta L_{\rm p}$.
Using the dynamic range $\drange=\gamma_2/\gamma_1$ as the remaining
free parameter, one finds:
\begin{eqnarray}
\gamma_1&=&\Gamma\left({q-2\over q-1}\right)
\left[{1-{\drange}^{1-q}\over 1-{\drange}^{2-q}}\right]
\end{eqnarray}
which depends only weakly on ${\drange}$ for $q>2$ and ${\drange}\gg1$.
Energy conservation implies
\begin{eqnarray}
Q_0&=& \gamma_1^{q-1} {\eta L_{\rm p}\over \Gamma mc^2}\; 
{(q-1) \over 1-{\drange}^{1-q}}\;.
\end{eqnarray}

%
\begin{figure}[htb]
\centerline{\psfig{figure=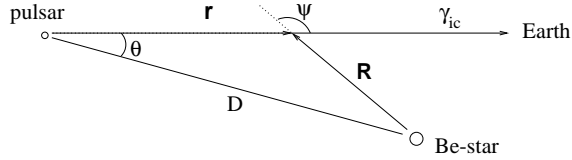,width=7.5cm}}
\caption{Sketch of the binary system defining angles and distances
used in the text.}
\label{geometry}
\end{figure}

For the purpose of calculating the inverse Compton emissivity of
an electron, we assume that the spectrum of the pulsar's companion star
can be approximated as monochromatic at frequency
$\nu_0=2.7 k_{\rm B}T_{\rm eff}/h$,
where $T_{\rm eff}$ is the effective temperature of the star
[Blumenthal \& Gould 1970].
We denote by
$n_\gamma(\epsilon,{\vec{R}},{\vec{\Omega}}) \,\diff{\vec{\Omega}} \, \diff \epsilon$
the differential number density of target photons moving within the solid angle
$\diff{\vec{\Omega}}$ of the unit vector ${\vec{\Omega}}$ at position
${\vec{R}}$ with respect to the star,
with energy between $E=\epsilon mc^2$ and $(\epsilon+\diff\epsilon)mc^2$
where $\epsilon$ is dimensionless and $m$ is the electron mass.
Then
\begin{eqnarray}
n_\gamma(\epsilon,{\vec{R}},{\vec{\Omega}})&=&
N(R) \, \delta(\epsilon-\epsilon_0) \, 
\delta({\vec{\Omega}}-{\hat{\vec{R}}})
\label{photdens}
\end{eqnarray}
where $N(R)=\Lstar/(4\pi R^2 c h\nu_0)$ is the photon density, 
$\epsilon_0=h\nu_0/mc^2$ is the dimensionless energy of the Be-star photons,
${\hat{\vec{R}}}={\vec{R}}/R$ is a unit vector,
and $\Lstar$ is the stellar luminosity.

The ratio of emission due to synchrotron losses to that due to
inverse Compton losses is, in the classical limit, determined
by the relative energy density in the magnetic field and in the
target photons.
It is therefore convenient to introduce the parameter
\begin{eqnarray}
b&\equiv&\sqrt{B^2/8\pi \over h\nu_0 N }
\;.
\label{bparam}
\end{eqnarray}

We assume that the pulsar-shock separation is small compared with the
pulsar-Be star separation $D$.
In this case, both the target photon density, $N \approx N(D)$,
and the angle between
the line of sight and the direction of the target photons, $\psi$,
remain constant over the shock surface.
The (isotropic) electron distribution function, integrated over the 
entire source can be easily derived from the kinetic equation
in two special cases.
If adiabatic losses dominate, the shape of the spectrum is unchanged
by the losses and one can approximate the distribution function as a
product of the loss time and the injection function:
\begin{eqnarray}
n_{\rm e}(\gamma) &=&
\Delta t \, Q_{\rm e}(\gamma)
\label{adiabdist}
\;.
\end{eqnarray}
If, on the other hand, radiative losses dominate, the kinetic equation
[e.g.\ Eq.~56 of Tavani \& Arons 1997]
is easily solved to give: 
\begin{eqnarray}
n_{\rm e}(\gamma)&=& {1\over 
\left<-\dot\gamma_{\rm ic}\right>+\left<-\dot\gamma_{\rm s}\right>}
\int_\gamma^\infty{\diff\gamma'\over 4\pi} \; Q_{\rm e}(\gamma')
\nonumber\\
&=& \left\{
\begin{array}{ll}
{\textstyle Q_0\over \textstyle 4\pi(q-1)} \;
{\textstyle \left( \gamma^{1-q} - \gamma_2^{1-q}\right)\over 
\textstyle (\left<-\dot\gamma_{\rm ic}\right>+
            \left<-\dot\gamma_{\rm s}\right>)} & 
{\rm for\ }\gamma_1<\gamma<\gamma_2 \;,\\
{\textstyle Q_0\over \textstyle 4\pi(q-1)} \;
{\textstyle \left( \gamma_1^{1-q} - \gamma_2^{1-q}\right)\over 
\textstyle (\left<-\dot\gamma_{\rm ic}\right>+
           \left<-\dot\gamma_{\rm s}\right>)} &
{\rm for\ }\gamma<\gamma_1\;.
\end{array}
\right.
\label{raddist}
\end{eqnarray}
The quantity $\left<\dot\gamma_{\rm ic}\right>$ is the rate of change
of the electron Lorentz factor due to inverse Compton losses,
averaged over scatterings and over an isotropic distribution of
electron directions (or equivalently, of incoming photon directions).
Similarly, $\left<\dot\gamma_{\rm s}\right>$ is the corresponding quantity
for synchrotron losses, averaged over pitch angle.
In deriving Eq.~(\ref{raddist}) we have used the continuous approximation
to the inverse Compton losses
[Blumenthal 1971] 
and ignored triplet pair production,
since the maximum value of the parameter 
$\gamma\epsilon_0$ which we consider is of order $500$
[Mastichiadis 1991].
The details of the calculations of $\left<\dot\gamma_{\rm ic}\right>$
and $\left<\dot\gamma_{\rm s}\right>$ are presented in Appendix A.

The rate at which the electrons emit inverse Compton photons
per steradian per second per energy interval depends on the angle
$\psi$ of Fig.~\ref{geometry} and is given by
\begin{eqnarray}
{\diff N_{\rm ic}\over \diff{\vec{\Omega}}\diff\varepsilon\diff t}&=&
\int\diff\gamma \; n_{\rm e}(\gamma)
\eta_{\rm ic}(\varepsilon,\gamma,\psi)
\label{dNic}
\end{eqnarray}
where $\eta_{\rm ic}(\varepsilon,\gamma,\psi)$ is the inverse Compton emissivity
-- the number of photons emitted per second per electron in unit energy interval
around $\varepsilon$.
Analogously, the corresponding synchrotron emission 
depends on the angle between the magnetic field in the source and the 
line of sight. This, however, is likely to vary considerably through the 
source, so that we average over all pitch-angles:
\begin{eqnarray}
{\diff N_{\rm s}\over \diff{\vec{\Omega}}\diff\varepsilon\diff t}&=&
\int\diff\gamma \; n_{\rm e}(\gamma) \;
\eta_{\rm s}(\varepsilon,\gamma)
\label{dNs}
\end{eqnarray}
where $\eta_{\rm s}(\varepsilon,\gamma)$ is the synchrotron emissivity.
The details of the calculation of the rates of emission
are presented in Appendix A.

Photons of sufficiently high energy produced by inverse Compton scattering,
may interact with the radiation field of the Be star on their way from the
source to the observer and create electron-positron pairs, 
$\gamma \gamma \rightarrow e^+e^-$.
If a substantial fraction of
the inverse Compton emission is absorbed, 
an electron positron pair cascade will result and the power will emerge at
lower photon energy in 
a beaming pattern which is likely to be reasonably isotropic. However,
this effect should not overwhelm the primary radiation from the shocked 
electrons close to the neutron star, since the reradiated power
occupies a larger solid angle than the intercepted power and the primary
radiation itself is in general more intense at lower photon energy. 
To a first approximation, therefore, we treat this effect as absorption.
The derivation of the effective optical depth due to pair-production,
$\tau(\varepsilon,\theta)$, is presented in Appendix~B.
The minimum energy required for pair creation via this process is
\begin{eqnarray}
\varepsilon_{\rm ic} \ge 2/[(1-\cos\psi)\epsilon]
\label{threshold}
\end{eqnarray}
which corresponds to roughly $51\,{\rm GeV}\times2/(1-\cos\psi)$
for the assumed monochromatic Be-star spectrum.

We denote by $F_E$ the total observed energy flux per unit energy interval
due to synchrotron and inverse Compton emission from the shocked, accelerated
electrons and positrons.
When corrected for the absorption by pair-production on Be-star photons,
one has
\begin{eqnarray}
E F_E = {\varepsilon^2 mc^2\over d^2}
\left({\diff N_{\rm s}\over \diff{\vec{\Omega}}\diff\varepsilon\diff t}
+{\diff N_{\rm ic}\over \diff{\vec{\Omega}}\diff\varepsilon\diff t}\right)
{\rm e}^{-\tau} \;,
\label{flux}
\end{eqnarray}
where $d$ is the distance to the emission region.

\section{Application to \PSR}

The published observations available to constrain the parameters
consist of a set of four detections at different epochs by the ASCA satellite
[Kaspi et al.\ 1995, Hirayama et al.\ 1996], 
a single set of measurements by the OSSE experiment 
[Grove et al.\ 1995]
taken over a range of epochs,
and upper limits from the COMPTEL and EGRET experiments
[Tavani et al.\ 1996].
In addition, the source has been detected at an epoch far from
periastron by the ROSAT satellite
[Cominsky, Roberts \& Johnston 1994],
but the measurement is not sufficiently precise to
constrain the model significantly, and will not be discussed here.

The emission in the range $2$ -- $10\,{\rm keV}$ 
(ASCA) varies with orbital phase close to periastron.
Of the four epochs measured, that closest to periastron stands out
from the other three because the flux is approximately a factor of two weaker
and the spectrum significantly softer. 
The sensitivity of the OSSE experiment is not
sufficient to detect such variability in the $50$ -- $200\,{\rm keV}$ band.
We therefore consider the OSSE observations as an average over epochs
spread over about 20~days around periastron. 

The most obvious constraint on the model is given by the very hard photon index
of roughly $-1.7$ measured by ASCA and OSSE and consistent with the relative
intensities of the fluxes in the two bands.
In the shock acceleration picture, the X-ray emission can be
interpreted as either inverse Compton or synchrotron emission.
However, the inverse Compton interpretation would require a
very low value for the Lorentz factor of the pulsar wind ($\Gamma\sim 500$),
compared to that derived for the Crab pulsar ($\Gamma\sim10^6$),
in order to satisfy the upper limits imposed by the EGRET measurements.
We do not consider the inverse Compton interpretation
of the X-ray emission further here.

The first fundamental question to be answered before attempting a fit
to the spectrum is whether the accelerated electrons are able to cool 
radiatively (by synchrotron emission or inverse Compton emission), or 
whether they first lose their energy by adiabatic losses
in the post-shock flow. 
Ball et al.\ [1998]
estimate a field of $\sim 1\,$G at the shock at periastron, in which case
electrons with a Lorentz factor of $10^6$ radiate synchrotron emission
in the X-ray band and have a synchrotron half life time of 
roughly 10 minutes.
A simple estimate of the timescale associated with the flow patterns --
and thus the adiabatic losses --
is that it is the stand-off distance of the pulsar wind shock, 
divided by the post-shock speed.
For an ultrarelativistic flow such as is thought to emerge from the pulsar,
the post-shock speed is $c/3$. On the other hand, the stand-off distance of the
pulsar wind shock is likely to scale with the distance between the stars.
We therefore approximate the timescale for adiabatic losses as
\begin{eqnarray}
\Delta t&=& f3D/c\;,
\label{adlosstime}
\end{eqnarray}
where $f\ll1$ to conform with our assumption that the emission region is always
situated close to the pulsar. Taking $f=0.1$ leads to a $\Delta t$ of roughly
$2\,$minutes close to periastron
and $\sim 22\,$minutes at apastron.
Thus it is by no means clear which loss mechanism
-- radiative or adiabatic -- dominates.
Since we do not attempt to model the flow pattern,
we present both possibilities,
bearing in mind that the real system might lie in an intermediate region.

Another fundamental question concerns the orbital variation of the
magnetic field strength in the radiating region.
The position of the pulsar wind shock is thought to be determined by a
balance between the ram pressures of the Be-star wind and the 
pulsar wind
[Melatos, Johnston \& Melrose 1995],
which might reasonably be expected to vary throughout the orbit.
To characterise the resulting variation in the magnetic field strength,
we adopt a simple scaling:
we assume that the parameter $b$ of Eq.~(\ref{bparam})
-- the ratio of the energy density in the magnetic field
to that in the Be-star's radiation field in the emission region -- is constant.
This is the case, for example, if the distance of the shock from both the
pulsar and the Be star is proportional to the separation of the two stars,
and if the magnetic field associated with the pulsar has a $1/r$ dependence.
Such a distance scaling in its turn results if the radial dependence of
the ram pressures of the two winds is the same (for example $1/r^2$).

\begin{table}[htb]
\caption{Parameters for the \PSR\ system.
References are:
1--Johnston et al.\ [1996];
2--Johnston et al.\ [1994];
3--Underhill \& Doazan [1982].}
\label{1259par}
\begin{tabular}{llcc}
{\bf Parameter} & ~ & {\bf Value} & {\bf Reference} \\
{\bf Pulsar}\\
period & $P$ & $47.762 \;{\rm ms}$  & 1\\
period derivative & $\dot{P}$ & $2.279\times 10^{-15}$ & 1 \\
surface magnetic field & $B_{\rm p}$ & $3.3\times 10^7\;{\rm T}$ & 1 \\
spin down luminosity & $L_{\rm p}$ & $8.3\times 10^{28}\;{\rm W}$ & 1 \\
{\bf Be star}\\
spectral type  & & B2 & 2 \\
effective temperature & $T_{\rm eff}$ & $2.28\times 10^4\; {\rm K}$ & 3 \\
radius & $R_*$ & $6.0 R_\odot = 4.2\times10^9\;{\rm m}$ & 3 \\
luminosity & $L_*$ & $8.8\times 10^3 L_\odot=3.3\times10^{30}\;{\rm W}$ & 3 \\
effective photon energy & $\epsilon_0$ &
         $2.7 k_{\rm B} T_{\rm eff}/ (m c^2) = 10^{-5}$ \\
mass & $M_*$ & $10 M_\odot=2\times10^{31}\;{\rm kg}$ & 3 \\
{\bf System}\\
eccentricity & $e$ & 0.87 & 1 \\
periastron separation & $D_\tau$ & $23 \Rstar=9.6\times10^{10}\;{\rm m}$\\
apastron separation & $D_a$ & $331 \Rstar=1.4\times10^{12}\;{\rm m}$\\
orbital inclination & $i$ & $35^\circ$ & 1 \\
orbital period & ~ & 1236.8 days & 2\\
distance & $d$ & 1.5 kpc & 1 
\end{tabular}
\end{table}

\subsection{Adiabatic losses}

The simplest case to discuss is that in which adiabatic losses rapidly quench
the synchrotron and inverse Compton emission.
Since adiabatic losses do not alter the slope of the electron distribution,
we can approximate their effect as one of switching 
off the synchrotron and inverse Compton radiation after a finite time interval 
$\Delta t$ which is short compared to a radiative cooling time. 
The electron density in the emission region is
then given by Eq.~(\ref{adiabdist}),
and the only orbital variation in the emitting distribution
enters through the adiabatic loss time (Eq.~\ref{adlosstime}),
in which we have set $f=0.1$.

%
\begin{figure}[htb]
\centerline{\psfig{figure=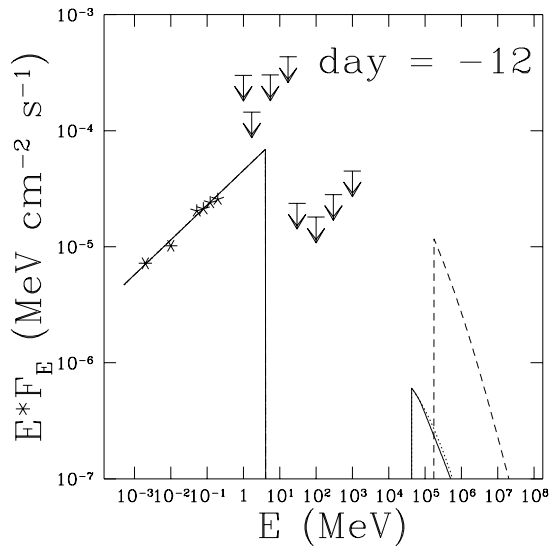,width=7.5cm}}
\caption{Model spectrum 12 days prior to periastron
when adiabatic losses dominate. 
The solid line shows the emission for $b=1$ ($B=3.2\,$G),
and the dotted line depicts the intrinsic emission,
uncorrected for photon-photon absorption. 
The dashed line shows the inverse Compton emission for $b=0.1$ ($B=\,0.32\,$G)
with injection rescaled to give the same synchrotron emission
as the higher field case.
Detections by ASCA and OSSE are shown as asterisks, 
and the upper limits are from COMPTEL and EGRET.}
\label{spectrum1}
\end{figure}

The photon index of $-1.7$ is matched in this case by injecting electrons
with a power-law index $q=2.4$, 
close to that which is thought to be produced in the Crab nebula.
In Fig.~{\ref{spectrum1}} we show a model spectrum computed for an epoch
12 days prior to periastron; the model parameters for this and subsequent
figures are given in Table \ref{modelpar}.
Also shown in this figure are the fluxes observed at the same epoch by ASCA, 
together with the observations by OSSE and the upper limits 
from the COMPTEL and EGRET experiments.
The sharp cut-off of the synchrotron emission above about $5\,$MeV and of 
the inverse Compton emission below about 
$50\,$GeV is an artifact of two of our approximations: 
that of sharp cut-offs at $\gamma_1$ and $\gamma_2$ in the
electron injection spectrum, and the \lq delta-function\rq\ approximation
to the emissivities. 
Replacing the delta-function emissivities with the correct forms
rounds off these cut-offs but does not change our conclusions.
Perhaps the most striking aspect of the model fit is the importance 
of the EGRET upper limits in constraining the upper extent
of the synchrotron emission.
One can immediately deduce that 
\begin{eqnarray}
10^{15}<\gamma_2^2 B<10^{17}
\end{eqnarray}
where $B$ is measured in gauss.
In addition, it can readily be seen that the electron injection function
must be close to a power law over at least a decade of $\gamma$
in order to fit the observations,
as stressed by
Grove et al.\ [1995].
\begin{table}[htb]
\caption{Parameters for the models presented in
Figs.\ \protect\ref{spectrum1}--\protect\ref{lcurve4}. $b$ is the square root
of the ratio of
magnetic to radiation energy density in the source, $q$ the power-law index of
the electron injection function, $\Gamma$ the Lorentz factor of the pulsar 
wind, $\eta$ the efficiency of conversion of spin-down power into injected
electrons and $\rho$ the range of Lorentz factors over which 
injection occurs, $\gamma_2$ being the upper cut-off.}
\label{modelpar}
\begin{center}
\begin{tabular}{lcccccccc}
{\bf Figure} & \multicolumn{5}{c}{\bf Model parameters} &
                \multicolumn{3}{c}{\bf ~} \\
~          & $b$ & $q$   & $\Gamma$        & $\eta$             & $\rho$ &Losses & ~~ & Implied $\gamma_2$  \\
2, 3       & 1   & $2.4$ & $5\times10^5$   & $4.4\times10^{-4}$ & $100$  & adiabatic & ~~ & $1.7\times10^7$ \\
2          & 0.1 & $2.4$ & $1.6\times10^6$ & $1.4\times10^{-2}$ & $100$  & adiabatic & ~~ & $5.4\times10^7$ \\
4, 5       & 1   & $1.2$ & $2\times10^6$   & $1.3\times10^{-2}$ & $100$  & radiative & ~~ & $1.2\times10^7$ \\
6, 7, 8    & 0.1 & $1.4$ & $5\times10^6$   & $1.5\times10^{-2}$ & $100$  & radiative & ~~ & $4.3\times10^7$ \\
\end{tabular}
\end{center}
\end{table}

Since in this case the electron distribution is not influenced by 
radiative losses, the synchrotron emission is unaffected by a change
in the magnetic field strength, provided the upper and lower cut-offs
of the injection spectrum ($\gamma_1$ and $\gamma_2$)
and the absolute level of emission (i.e., $\eta$) are appropriately rescaled.
The resulting inverse Compton emission, on the other hand, 
is strongly affected by the choice of $b$.
In Fig.~\ref{spectrum1}, the solid line shows the emission 
12 days prior to periastron, computed under the assumption that the
energy density in the magnetic field equals that in Be-star photons
at the position of the pulsar, i.e., $b=1$,
which corresponds to a field strength of $3.2\,$G at this epoch.
The substantially larger inverse Compton flux depicted by the
dashed line is that which results from a model which produces the same 
synchrotron emission but has $b=0.1$, corresponding to $B=0.32\,$G.
The inverse Compton emission increases by considerably
less than a factor $b^{-2}$ because of the Klein-Nishina effects.

At energies above about $50\,$GeV, the spectrum is affected by
absorption due to photon-photon pair production in the
radiation field of the Be star.
We show this explicitly in Fig.~\ref{spectrum1} by plotting the 
intrinsic emission, without allowance for absorption, as a dotted line. 

%
\begin{figure}[htb]
\centerline{\psfig{figure=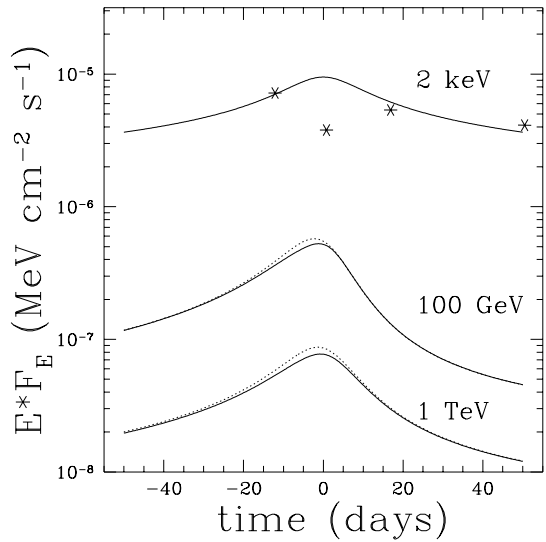,width=7.5cm}}
\caption{Model light curves in the X-ray and $\gamma$-ray bands
around periastron when adiabatic losses dominate. 
The abscissa gives the time in days relative to periastron.
The solid lines show the emission for $b=1$, for the same parameters as 
in Fig. \protect\ref{spectrum1}, 
and the dotted lines depict the intrinsic emission,
uncorrected for photon-photon absorption.}
\label{lcurve1}
\end{figure}

The light curves in the X-ray and hard 
$\gamma$-ray bands, when adiabatic losses dominate,
are depicted in Fig.~\ref{lcurve1}.
There is no emission between about $5\,$MeV and $50\,$GeV in this case
for two reasons:
firstly, because there is a lower limit to the energy at which electrons
are injected, and secondly, our crude treatment of adiabatic losses
simply switches off emission after a finite time and does not model
the cooling of electrons to energies below the lower injection cut-off. 
The observations by ASCA are indicated by asterisks and are to
be compared to the $2\,$keV light curve.
The increase in the model light curve at $2\,$keV is
clearly contradicted by the observations.
This point has been discussed in detail by
Tavani \& Arons [1997],
and we return to it in the following section.
The behaviour of the model is the result of two competing effects.
Firstly, the assumption of constant $b$ implies that the magnetic field
in the emission region is inversely proportional to the separation of the stars.
For a fixed power-law electron distribution, 
the emission at a given frequency increases according to $B^{(q+1)/2}$. 
Folding this with the assumed dependence of $\Delta t$, we find the emission
varies as $D^{(1-q)/2}$.
With $q=2.4$, the increase in emission between 
$-12\,$days and periastron is about a factor of 1.3.

Although the X-ray light curve does not fit the observations,
the inverse Compton light curves display interesting features which persist in the
more complicated cases in which radiative cooling dominates.
Since $b$ is assumed constant over the orbit,
the ratio of inverse Compton to synchrotron luminosity is approximately constant.
However, the inverse Compton radiation is anisotropic,
being stronger when the target photons are scattered head-on into
the line of sight, than when they are deflected by an angle of
less than $90^\circ$.
The scattering angle $\psi$ is a function of orbital phase,
and, because of the inclination of the orbit,
it is larger before periastron than after.
This accounts for the asymmetry of the flux evolution at $100\,$GeV
with respect to periastron.
The $100\,$GeV emission is systematically stronger before periastron,
reaching a maximum close to the point at which the scattering angle
peaks, at $-6.4\,$days.
The same behaviour is displayed by the intrinsic
(unabsorbed) emission at $1\,$TeV (dotted line). 
However, the largest scattering angle is achieved when the pulsar is
almost \lq behind\rq\ the star, at which point the optical depth along
the line of sight due to absorption by photon-photon pair production is greatest.
This effect reduces the emerging flux at $1\,$TeV just prior to periastron.
Nevertheless, its value $20\,$days before periastron is
about a factor of 1.5 larger than $20\,$days after periastron.

\subsection{Radiative losses}

When radiative losses dominate over adiabatic losses,
the shape of the electron distribution function varies
with orbital phase,
because the loss terms $\left<\dot\gamma_{\rm ic,s}\right>$
are functions of $\gamma$ and depend on the stellar separation. 
Assuming that the timescale associated with the orbital motion
is much longer than the radiative loss timescale,
the instantaneous values of the loss terms can be inserted into
Eq.~(\ref{raddist}) to find the electron distribution.

%
\begin{figure}[htb]
\centerline{\psfig{figure=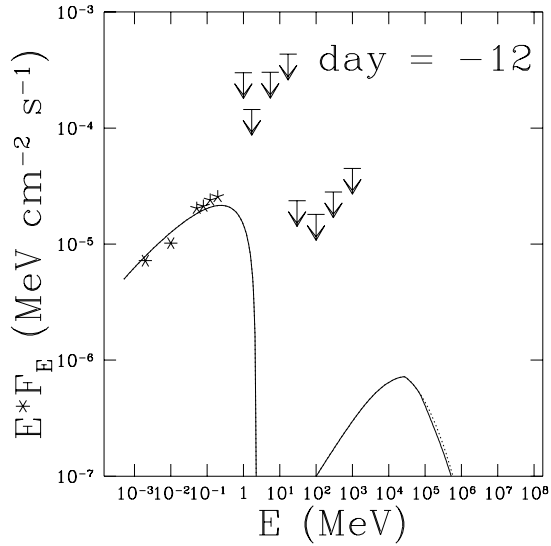,width=7.5cm}}
\caption{Model spectrum 12 days prior to periastron when
radiative (synchrotron) losses dominate. 
The solid line shows the emission for $b=1$ ($B=3.2\,$G),
and the dotted line depicts the intrinsic emission,
uncorrected for photon-photon absorption. 
Detections by ASCA and OSSE are shown as asterisks, 
and the upper limits are from COMPTEL and EGRET.}
\label{spectrum2}
\end{figure}

The synchrotron loss timescale is inversely proportional to the
electron Lorentz factor $\gamma$ throughout the entire spectrum.
Consequently, when synchrotron losses dominate, the
observed (photon) spectral index is $-(q+2)/2$.
The photon index of $-1.7$ measured by ASCA and OSSE can therefore
be reproduced by choosing an injection spectrum with $q=1.4$.
However, such a hard injection spectrum displays a much broader
\lq roll-over\rq\ in the neighbourhood of the upper cut-off,
as can be seen from Eq.~(\ref{raddist}).
Compensating for this with an even harder injection spectrum
gives a better fit to the OSSE observations, as is shown
in Fig.~\ref{spectrum2}.
Here we have chosen $b=1$, so that, because the Klein-Nishina effects
weaken the inverse Compton losses, synchrotron cooling dominates.
The synchrotron emission is similar to that of Fig.~\ref{spectrum1},
except for the smooth roll-over. 
However, the inverse Compton emission contains a new, rather weak,
component in the EGRET waveband.
This arises from electrons which have cooled to Lorentz factors
lower than the minimum at which injection took place, $\gamma_1$. 
The photon index in this band is $-1.5$,
appropriate to cooling electrons which were all injected at
higher energy.

%
\begin{figure}[htb]
\centerline{\psfig{figure=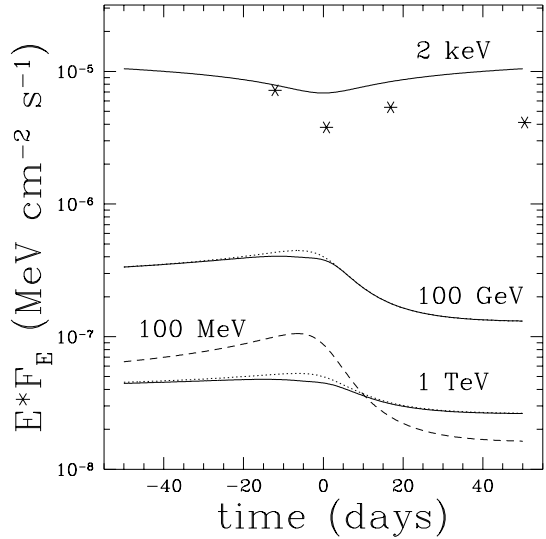,width=7.5cm}}
\caption{Model light curves in the X-ray and $\gamma$-ray bands around
periastron when radiative (synchrotron) losses dominate. 
The solid lines show the emission for $b=1$,
for the same parameters as in Fig. \protect\ref{spectrum2}. 
The dotted lines depict the intrinsic emission,
uncorrected for photon-photon absorption.
Note that the emission at $100\,$MeV was not present
in the adiabatically cooled case (Fig.~\protect\ref{spectrum1}).}
\label{lcurve2}
\end{figure}


The corresponding light curves are shown in Fig.~\ref{lcurve2}.
The light curve at $2\,$keV now displays a slight dip at periastron,
in contrast to the case where adiabatic losses
dominate shown in Fig.~\ref{lcurve1}.
This demonstrates the changes in the shape of the electron distribution
with orbital phase.
The total emitted luminosity is independent of orbital phase because
the efficiency $\eta$ of conversion of pulsar spin-down luminosity to
relativistic electrons is held constant, and because essentially all
of the energy in the electrons is radiated.
The relative importance of synchrotron and inverse Compton emission is controlled
by the parameter $b$, which is also constant, so that the synchrotron
luminosity does not vary with phase.
However, since the magnetic field in the source region is stronger
at periastron, electrons radiate at higher frequencies there,
i.e., the entire spectrum in the $E F_E$ plot shown in
Fig.~\ref{spectrum2} is shifted horizontally to the right.
The size of the effect this has on the light curve at $2\,$keV
depends on the spectral index.
The observed dip by roughly a factor of 2 would require a change
of one order of magnitude in the magnetic field, given the observed
photon index of $-1.7$.
In our model the magnetic field increases by only a factor of 
$1.5$ between the epoch $-12\,$days and periastron,
giving a change of only 14\% in the flux.

As in the case when adiabatic losses dominate, the $\gamma$-ray emission is
sensitive to a change in the parameter $b$. 
The inverse Compton loss timescale is inversely proportional to
$\gamma$ for very low Lorentz factors $\gamma\epsilon_0\ll 1$.
However, unlike the synchrotron case,
Klein-Nishina corrections weaken the losses at higher $\gamma$ until,
at about $\gamma\epsilon_0\approx 1$
(corresponding to $\gamma\approx 10^{5}$ here),
the timescale becomes independent of $\gamma$.
This behaviour significantly complicates the resulting spectrum,
since for electrons radiating synchrotron emission in the X-ray band,
$\gamma\epsilon_0 > 1$.
Where inverse Compton losses dominate, the electron distribution tends
to have the same power law index as the injection spectrum,
and the photon index of synchrotron emission from these electrons
is thus $\sim0.5$ harder than that produced where synchrotron cooling dominates.
As the magnetic field is reduced,
inverse Compton cooling first starts to dominate at the low frequency end
of the spectrum since the weakening due to Klein-Nishina effects is
not strongly pronounced there.
This tends to give a harder spectrum in the ASCA range than in the OSSE range. 

%
\begin{figure}[htb]
\centerline{\psfig{figure=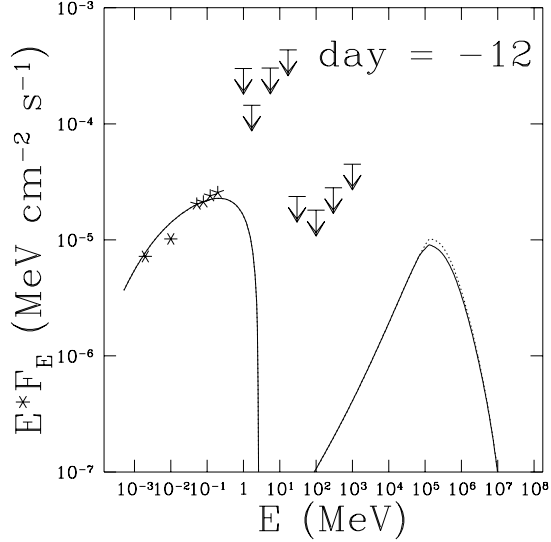,width=7.5cm}}
\caption{Model spectrum 12 days prior to periastron 
when radiative (inverse Compton) losses dominate. 
The solid lines show the emission for $b=0.1$ ($B=0.32\,$G),
and the dotted line depicts the intrinsic emission,
uncorrected for photon-photon absorption.
Note the curvature of the spectrum at energies between
$1-10\,$keV which results from inverse Compton losses.}
\label{spectrum3}
\end{figure}

\begin{figure}[hbt]
\centerline{\psfig{figure=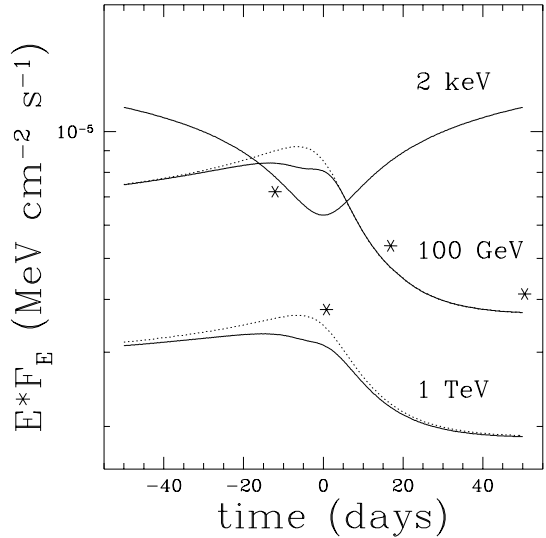,width=7.5cm}}
\caption{Model light curves in the X-ray and $\gamma$-ray bands
around periastron when radiative (inverse Compton) losses dominate. 
The solid lines show the emission for $b=0.1$,
for the same parameters as in Fig. \protect\ref{spectrum3}. 
The dotted lines depict the intrinsic emission,
uncorrected for photon-photon absorption.}
\label{lcurve3}
\end{figure}

An example of this effect is shown in Fig.~\ref{spectrum3},
where we have chosen the parameter $b=0.1$.
As in the case where adiabatic losses dominate,
the inverse Compton flux level is higher than for the stronger field case
shown in
Fig.~\ref{spectrum2}, and is now comparable to that of the synchrotron emission.
However, the synchrotron spectrum now has a pronounced turnover
at low energies and
its shape is quite different
to that shown in Fig.~\ref{spectrum2}.
In order to compensate for the hardening effect of the inverse Compton cooling,
the fit presented in Fig.~\ref{spectrum3} used an injection spectrum
with $q=1.4$, softer than the injection spectrum with
$q=1.2$ used for Fig.~\ref{spectrum2}.
The harder model 
spectrum in the ASCA range has a marked effect on the light curve, 
which is shown in Fig.~\ref{lcurve3}.
The flux at $2\,$keV is now much more sensitive to a shift of the
emission to higher frequency, and therefore the dip in flux at periastron
is more pronounced.

%

Despite the strong difference in the behaviour of the X-ray light curve
in the cases where adiabatic and radiative losses dominate,
the characteristics of the $\gamma$-ray curves are essentially the same.
This is due in the first place to the effect of the scattering angle,
which has the same orbital dependence in both cases.
In the second place, the photon index in the $\gamma$-ray range is much softer
when radiative losses dominate, and this tends to enhance the emission
when the spectrum is boosted to higher frequencies,
whereas it decreases the emission at $2\,$keV.
In the adiabatic case, where the luminosity is not fixed,
the increased photon density at periastron causes a rise in the
inverse Compton flux,
whilst at the same time the stronger magnetic field causes
a rise in the synchrotron flux (see Fig.~\ref{lcurve1}).

\begin{figure}[htb]
\centerline{\psfig{figure=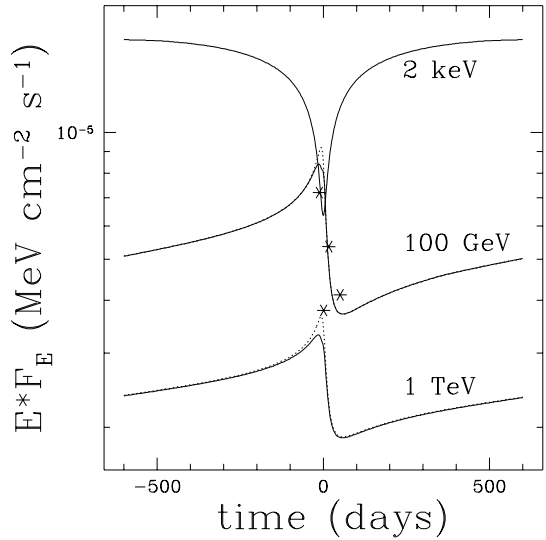,width=7.5cm}}
\caption{The model light curves of Fig.~\protect\ref{lcurve3}
extended to cover the entire orbit.}
\label{lcurve4}
\end{figure}

Fig.~\ref{lcurve4} shows the light curve of Fig.~\ref{lcurve3} extended 
over the entire orbit.
According to our model, X-ray emission should persist at all epochs.
This is in agreement with the ROSAT detection near apastron
[Cominsky, Roberts \& Johnston 1994],
although the observed flux was rather lower than the model flux.
The most prominent feature of the light curves is the 
steady increase in $\gamma$-ray emission from soon after one periastron
to just prior to the next periastron.
This mirrors the dependence of the 
inverse Compton radiation mechanism on the angle between the
line of sight and the vector connecting the stars.

%

\section{Discussion}

In presenting models of the high energy emission from shock accelerated
electrons in \PSR\ we have used simple scaling rules for those quantities
which depend on the flow pattern of plasma in this binary system.
Thus, the stand-off distance of the pulsar wind termination shock 
is the basic length scale governing
the importance of adiabatic losses, 
and we assume that the ratio of 
this length to the stellar separation is constant.
It is clear from the unpulsed radio emission detected from the
system near periastron that the pulsar interacts with a disk around the
Be star
[Ball et al.\ 1998].
Nevertheless, we have assumed that the position of the termination shock 
is not affected by the disk, and have also taken the magnetic field strength
at the shock front as well as the density in target photons from the Be star
to be unchanged by the presence of the disk.
In this case a $1/r$ dependence of the magnetic field 
together with the $1/R^2$ dependence of the photon density 
(see Fig.\ \ref{geometry}) ensures that the ratio of the 
energy density in the magnetic field to that in the 
target photons is independent of orbital phase. 

These approximations are plausible except when the pulsar is very
close to periastron.
Changes in the observed pulsed radio emission
[Johnston et al.\ 1996]
and observations and interpretation of the unpulsed radio emission
seen around periastron
[Johnston et al.\ 1998,
Ball et al.\ 1998],
imply that the interaction of the
pulsar wind and the Be-star disk is confined to the period
from about day $-25$ to $+25$.
There are few observational data of X-ray and $\gamma$-ray emission
available at epochs outside this range.
We have therefore made comparisons of our model predictions
with observations taken close to periastron.
The models presented indicate the trends produced by effects such
as the competition between synchrotron and inverse Compton cooling.
In particular, our results indicate that the influence of inverse Compton
cooling on the spectral index in the X-ray band is strongly affected by the
functional dependence of the Klein-Nishina corrections on the electron
Lorentz factor.
However, they cannot be expected to accurately fit the observations
made close to periastron.

In common with most discussions of particle acceleration at pulsar 
wind termination shocks
[e.g.\ Kennel \& Coroniti 1984],
we assume that the mechanism produces a featureless power-law 
distribution between two sharp cut-offs.
In reality, the upper cut-off is likely to be determined by competition
between radiative losses and acceleration,
which may set in fairly sharply in energy space if the acceleration rate 
is a decreasing function of Lorentz factor and losses are increasing. 
The lower cut-off, on the other hand, depends on 
the mechanism by which some electrons from the relativistic MHD wind 
are decelerated, and the physical basis for the assumption of a
sharp lower cut-off is unclear.
These uncertainties should be kept in mind when judging
the results of detailed spectral modelling.

All of the models that we have presented require that only a small
fraction ($\sim1\%$) of the pulsar spin down luminosity be radiated
by the shocked electrons and positrons of the pulsar wind.
The efficiencies are comparable to those inferred for
the Crab nebula
[De Jager \& Harding 1992]
and are in agreement with those suggested by Tavani \& Arons [1997].
They are consistent with the likelihood that ions carry
much of the wind luminosity and that a large fraction of the power
is used to drive the expansion of the nebula.
The value of $\epsilon_0$ that we have used,
together with the high Lorentz factor of the unshocked pulsar wind,
means that the inverse Compton scattering is well into the
Klein-Nishina regime.
If the energy of the target photons is somewhat lower,
then the absorption due to pair production would be larger
(albeit above a higher threshold energy)
and the orbital variation of the hard $\gamma$-ray flux would be larger
than shown in our model light curves.

The problem of fitting the X-ray observations close to periastron 
has been thoroughly discussed by
Tavani \& Arons [1997],
and the properties of the flow pattern, not considered in our work,
play a crucial role in their analysis.
The reduction of the flux at $2\,$keV and simultaneous spectral softening
at periastron seen by ASCA is reproduced by taking account of the
competition between adiabatic and radiative losses
and by adopting strong shielding (or shadowing) of the emission region from 
the target photons whilst the pulsar is traversing the Be-star disk, 
(assumed to be whenever the orbital phase is outside the range
$-50^\circ<\phi<50^\circ$). 
The strong inverse Compton cooling which sets in upon emergence into
the full photon flux within a few days of periastron is responsible
for both transferring emission out of the X-ray band into the
$\gamma$-ray region and softening the electron distribution. 
The details of the conclusions reached by
Tavani \& Arons [1997]
may be modified by the dependence of the Klein-Nishina corrections
on the electron energy which we have shown to be important.
Calculations of models which include both the flow pattern and the
full functional form of the Klein-Nishina corrections are warranted.

Our results show that the $\gamma$-ray emission is not sensitive to the
modelling of the X-ray flux close to periastron.
The major parameter governing the expected $\gamma$-ray 
intensity is the magnetic field strength,
and the values we have adopted are relevant to the \PSR\ system
[Ball et al.\ 1998].
The spectral index in the $100\,$GeV to TeV 
range is generally much softer than in X-rays because of the fall-off of the 
Klein-Nishina cross section at high energy, but the 
fluxes we predict are of the order of 
$10^{-10}\,{\rm photons\,cm^{-2}\,s^{-1}}$ at $100\,$GeV -- 
well above the sensitivity threshold of proposed Cerenkov imaging telescopes
[Kawachi 1997].
The light curve of these $\gamma$-rays is affected in the TeV range 
by absorption on Be-star photons, but at lower energies is
dominated by the change of scattering angle through the orbit,
an effect which should appear whenever the pulsar is out of the
shadow of the Be-star disk, i.e., over almost the entire orbit.
Other production mechanisms, such as the decay of $\pi^0$ particles
produced in nuclear collisions, or the inverse Compton scattering
of synchrotron photons or of photons from the microwave background,
are expected to give isotropic emission.
Any variation of the flux produced by these mechanisms should
then be symmetric about periastron.
Detection of an asymmetric variation, such as we predict,
of the hard $\gamma$-ray flux from \PSR\ would provide conclusive proof
that inverse Compton scattering of the Be-star photons is the
responsible radiation mechanism.

\begin{ack}
%
We thank Simon Johnston for finding this pulsar,
and for many helpful discussions. Our collaboration was
made possible by support for
J.K. from the RCfTA, University of Sydney,
under its international visitor programme and for
L.B. from the MPIK Heidelberg.
O.S. thanks the RCfTA, University of Sydney,
for its support during work on this project,
and the Norwegian Research Council
for support of this work through a postgraduate grant.
\end{ack}

\section*{Appendix A: Inverse Compton and\protect\\synchrotron emissivities}
An electron of Lorentz factor $\gamma$ scatters target photons at an average 
rate 
\begin{eqnarray}
\dot N_{\rm ic}&=&
c\int \diff{\vec{\Omega}}\int\diff \epsilon\,
{\epsilon'\over\gamma\epsilon} \; n_\gamma(\epsilon,{\vec{R}},{\vec{\Omega}})
\sigma_{\rm KN}(\epsilon')
\label{scrate}
\end{eqnarray}
[Jones 1965],
where $\epsilon$ is the energy of the target photon,
$\epsilon'$ is the energy of the target photon seen in the
electron rest frame and 
$\sigma_{\rm KN}(x)$
%
%
is the Klein-Nishina cross section.
On inserting (\ref{photdens}) into Eq.~(\ref{scrate}) one finds
\begin{eqnarray}
\dot N_{\rm ic}&=&c N(R) \; \sigma_{\rm KN}(\epsilon') \; 
{\epsilon'\over\gamma\epsilon_0} \,.
\end{eqnarray}
For an electron moving at an angle $\psi$ to ${\vec{R}}$, one has  
\begin{eqnarray}
\epsilon'=\gamma\epsilon_0(1-\beta\cos\psi)
\label{epsprime}
\end{eqnarray}
where $c\beta$ is the speed of the electron,
and thus the scattering rate depends implicitly on the 
direction of motion of the electron relative to that of the photon,
and so $\dot N_{\rm ic}(\gamma,\psi)$.

The rate of change of electron energy averaged over 
scatterings remains a function of the direction of motion, 
and is given by
[Jones 1965],
\begin{eqnarray}
\dot\gamma_{\rm ic}(\gamma,\psi)&=&-{3\over8} \; c N(R) \; \sigmaT \; 
{{\epsilon'}^2\over\epsilon_0}
\left(1-{\epsilon_0\over\epsilon'\gamma}-
{\epsilon_0\over\gamma}\right)F_{\rm loss}(\epsilon')
\label{elossrate}
\end{eqnarray}
where $\sigmaT$ is the Thomson cross section and 
\begin{eqnarray}
F_{\rm loss}(x)&=&\int_0^2 \diff f\,
{f(f^2-2f+2)\over (1+xf)^3}\left[
1+{(xf)^2\over(f^2-2f+2)(1+xf)}\right]
\label{Floss}
\end{eqnarray}
which can be integrated by elementary techniques, to give
\begin{eqnarray}
F_{\rm loss}(x)&=&
{-2x(10x^4-51x^3-93x^2-51x-9)
\over
{3x^4(1+2x)^3}}
\nonumber\\
&&+{(x^2-2x-3)\ln(2x+1) \over x^4} \,.
\label{Flossa1}
\end{eqnarray}
For small $x$ this
can be expanded as a power series giving
\begin{eqnarray}
F_{\rm loss}(x)
\approx {8\over 3}-{56\over 5}x+{196\over 5}x^2-{12928\over 105}x^3\ldots
\hspace*{1cm} x\ll 1\;,
\label{Flossa2}
\end{eqnarray}
which is useful for numerical calculation.

We denote the average of $\dot\gamma_{\rm ic}$ over an isotropic distribution
of electron directions (or, equivalently, incoming photon directions) by
\begin{eqnarray}
\left<\dot \gamma_{\rm ic}\right> =
{1\over 2} \int_{-1}^1 {\rm d}\!\cos\psi \; \dot \gamma_{\rm ic}
\label{gdotic}
\end{eqnarray}
which it is most convenient to evaluate numerically using
equations (\ref{Flossa1}) and (\ref{Flossa2}).
Alternatively, it can be written in the form
\begin{eqnarray}
\left<\dot \gamma_{\rm ic}\right>&=&
-{4\over3} \; c N\sigmaT \; \gamma^2\epsilon_0 \; G(\gamma,\epsilon_0)
\end{eqnarray}
where, after substituting Eq.~(\ref{Floss}) into Eq.~(\ref{gdotic})
and reversing the order of the integrals over $\cos\psi$ and $f$,
it can be shown that
\begin{eqnarray}
G(\gamma,\epsilon)&=&
{9\over 64\gamma^4\epsilon^3\beta}\{
\gamma[f_1(\epsilon\bar{\gamma})-f_1(\epsilon/\bar{\gamma})]
-\epsilon[f_2(\epsilon\bar{\gamma})-f_2(\epsilon/\bar{\gamma})]\}
\end{eqnarray}
with $\bar\gamma=\gamma(1+\beta)=\gamma+(\gamma^2-1)^{1/2}$
and
\begin{eqnarray}
f_1(z)&=&
(z+6+3/z)\ln(1+2z)
-(22z^3/3+24z^2+18z+4)(1+2z)^{-2}\nonumber\\&&
-2
+2{\rm Li}_2(-2z)
\nonumber\\
f_2(z)&=&
(z+31/6+5/z+3/2z^2)\ln(1+2z)
-(22z^3/3+28z^2+103z/3\nonumber\\&&+17+3/z)(1+2z)^{-2}
-2
+2{\rm Li}_2(-2z)
\end{eqnarray}
where the function ${\rm Li}_2(z)$ is the Eulerian dilogarithm.
This result was first derived by
Jones [1965].

In the inverse Compton scattering regime,
in which the energy of the incoming photon is negligible,
the average energy of the scattered photon is
\begin{eqnarray}
\bar\varepsilon_{\rm ic}(\gamma,\psi)&=&{-\dot\gamma_{\rm ic}\over
\dot N_{\rm ic}} \;.
\label{avenergy}
\end{eqnarray}
This photon is emitted in the direction of motion of the electron,
and the emissivity -- the number of photons emitted per second per electron
in unit energy interval around $\varepsilon$ -- can be approximated by
\begin{eqnarray}
\eta_{\rm ic}
(\varepsilon,\gamma,\psi)&=&
\dot N_{\rm ic}(\gamma,\psi) \, \delta(\varepsilon-\bar\varepsilon_{\rm ic}(\gamma,\psi))
\;,
\label{etaic}
\end{eqnarray}
[Felten \& Morrison 1966].

Synchrotron radiation can be treated in an analogous manner
[Hoyle 1960],
using the 
\lq delta-function\rq\ approximation in which it is assumed that all
the emission from a given electron occurs at the energy
$\bar\varepsilon_{\rm s}(\gamma)$.
The rate of change of the electron Lorentz factor due to synchrotron
losses, averaged over pitch angle, is
\begin{eqnarray}
\left<\dot\gamma_{\rm s}\right>&=&-{4\over3} \; c N(R) \; 
\sigmaT\gamma^2\epsilon_0 b^2
\label{gammas}
\end{eqnarray}
and the pitch-angle averaged synchrotron emissivity per electron becomes
\begin{eqnarray}
\eta_{\rm s}(\varepsilon,\gamma)&=& 
\dot N_{\rm s}
\delta(\varepsilon-\bar\varepsilon_{\rm s}(\gamma))
\nonumber\\
&=&
{\left<-\dot\gamma_{\rm s}\right>\over \bar\varepsilon_{\rm s}(\gamma)} \; 
\delta(\varepsilon-\bar\varepsilon_{\rm s}(\gamma)) \;.
\label{etas}
\end{eqnarray}
The average frequency of the synchrotron photon is given by
\begin{eqnarray}
\bar\varepsilon_{\rm s}(\gamma)&=&
\gamma^2 b \sqrt{{\pi^3\over 8} \; {\epsilon_0\,N r_0^3\over\alpha_{\rm f}^2}}
\label{synchfreq}
\end{eqnarray}
where $\alpha_{\rm f}$ is the fine-structure constant and $r_0$
the classical radius of the electron.

Substitution of the delta-function approximation (\ref{etaic})
for the inverse Compton emissivity into Eq.~(\ref{dNic})
implies that the rate of emission of inverse Compton photons
per steradian per second per energy interval is
\begin{eqnarray}
{\diff N_{\rm ic}\over \diff{\vec{\Omega}}\diff\varepsilon\diff t}&=&
n_{\rm e}(\gamma_{\rm ic}) \;
\dot N_{\rm ic}(\gamma_{\rm ic},\psi)
\; {{\rm d}\gamma_{\rm ic}\over {\rm d}\varepsilon} \,.
\end{eqnarray}
The Lorentz factor $\gamma_{\rm ic}(\varepsilon,\psi)$
of an electron which emits inverse Compton photons 
of energy $\varepsilon$ is defined by the implicit relation
$\varepsilon=\bar\varepsilon_{\rm ic}(\gamma_{\rm ic},\psi)$. 

Similarly, from (\ref{gammas}) and Eq.~(\ref{dNs}),
the rate of emission of synchrotron photons
per steradian per second per energy interval is
\begin{eqnarray}
{\diff N_{\rm s}\over \diff{\vec{\Omega}}\diff\varepsilon\diff t}&=&
n_{\rm e}(\gamma_{\rm s}) \;
\dot N_{\rm s}(\gamma_{\rm s})
\;{{\rm d}\gamma_{\rm s}\over {\rm d}\varepsilon}
\end{eqnarray}
where now the Lorentz factor $\gamma_{\rm s}$ 
of an electron which emits synchrotron radiation at energy $\varepsilon$
is defined by $\varepsilon=\bar\varepsilon_{\rm s}(\gamma_{\rm s})$, 
and from Eq.~(\ref{synchfreq}), 
\begin{eqnarray}
\gamma_{\rm s}&=& \sqrt{\varepsilon\over b}
\left[{\pi^3 \over 8}\;
{\epsilon_0 N_0 r_0^3 \over \alpha_{\rm f}^2}\right]^{-1/4}\;.
\label{gammas2}
\end{eqnarray}


\section*{Appendix B: Opacity due to photon-photon pair production}

The computation of the optical depth due to photon-photon pair production
is most conveniently treated in the zero momentum frame,
commonly referred to as the centre of momentum 
frame (CM frame) of the photon-pair
[Jauch \& Rohrlich 1976, Svensson 1982]. 
In the CM frame, the cross section is
\begin{equation}
\bar\sigma = \frac{\pi r_0^2}{2} (1-\zeta^2) 
\left[(3-\zeta^4) \ln{\left(\frac{1+\zeta}{1-\zeta}\right)}-
2\zeta (2-\zeta^2)\right],
\end{equation}
where $\zeta=\sqrt{\bar\epsilon^{\,2}-1}/\bar\epsilon$ and 
bars denote quantities measured in the CM frame in which
each photon has an energy $\bar\epsilon$. 
The photon energy in the CM frame is related to the photon energies in the
observer's frame through the invariant
\begin{equation}
E^2-c^2 P^2 = 2 (1-\cos{\psi}) \varepsilon_{\rm ic}\epsilon
= 4\bar\epsilon^{\,2},
\end{equation}
where $E$ is the total energy and $P$ is the magnitude of ${\vec{P}}$,
the total momentum of the photon pair,
and $\psi$ is now the angle between the momentum vectors of the
high-energy photon and the target photon in the observer's frame.
The threshold condition is $\bar\epsilon>1$.

The optical depth between two points on a photon trajectory
is a Lorentz invariant, and is given by
\begin{eqnarray}
\diff\tau &=&  
2\diff\bar r
\int\,\diff\bar\epsilon \,\diff{\bar{\vec{\Omega}}} \, \bar\sigma\,\bar n_\gamma  \, 
=\, 
2\diff\bar r\int \,
\diff\epsilon \, \diff{\vec{\Omega}} \, \bar\sigma\,
\frac{\bar\epsilon}{\epsilon}\,n_\gamma  \, ,
\label{eqn:depth}
\end{eqnarray}
where $\diff\bar r$ is the spatial separation of the points in the CM frame.
The factor $2$ depends on the convention adopted for the definition of the 
cross section and stems from the fact that the relative speed of the
photons in the CM frame is $2c$
[see Jauch \& Rohrlich 1976].
The conversion from quantities in the CM frame to those in the observer's frame
in (\ref{eqn:depth}) exploits the Lorentz invariance of $n_\gamma/\epsilon^2$ and
$\epsilon \diff{\vec{\Omega}} \diff\epsilon$. 

The line element in the CM frame is given by
$\diff \bar r = \diff r \gamma_{\rm b} (1-\beta_{\rm b} \cos{\chi})$, 
where $\gamma_{\rm b}=(1-\beta_{\rm b}^2)^{-1/2}$ is the Lorentz factor
of the boost from the observer's
frame to the CM frame and $\chi$ is the angle
between the Lorentz boost and the ray path.
Assuming monoenergetic target photons from the Be star (Eq.~\ref{photdens}),
it follows that
\begin{eqnarray}
\diff\tau &=& N(R) \; \bar\sigma(\bar\epsilon) \; (1-\cos{\psi}) \diff r,
\label{eqn:depth3}
\end{eqnarray}
where $N(R)$ is the number density of target photons at a distance $R$ from the
Be star (see Fig.~\ref{geometry}).

There is a one-to-one correspondence between $r$
along a given ray (i.e.\ for a given angle $\theta$) and the
energy, $\bar\epsilon$, of the photons in the CM frame.
The optical depth from the pulsar to the Earth, given by (\ref{eqn:depth}),
may therefore be rewritten as an integral over the
CM-frame photon energy instead of the coordinate along the ray $r$.
Substituting $\alpha=\bar\epsilon/\sqrt{\varepsilon_{\rm ic}\epsilon_0}$
one finds 
\begin{equation}
\tau(\varepsilon_{\rm ic},\theta) = \frac{4N_0 D}{\sin{\theta}} 
\int_{1/\sqrt{\varepsilon_{\rm ic}\epsilon_0}}^{\sqrt{(1+\cos{\theta})/2}}
{\rm d}\alpha \; \frac{\alpha^2}{\sqrt{1-\alpha^2}}
\;\bar\sigma(\alpha\sqrt{\varepsilon_{\rm ic}\epsilon_0}) \;.
\label{eqn:depth4} 
\end{equation}
The upper limit of the integral in (\ref{eqn:depth4})
corresponds to the position of the pulsar.
Pair production is only possible if the energy of the photon 
exceeds the threshold Eq.~(\ref{threshold}).
As $\psi$ becomes small at a large distances from the star,
pair production ceases at some finite point.
The lower limit of the integral in (\ref{eqn:depth4})
corresponds to this cut-off.

{}
\end{document}